\documentclass{mem}
\usepackage{natbib}\usepackage{txfonts}\usepackage{balance}
\usepackage{flushend}
\usepackage{graphicx}
\usepackage[a4paper,breaklinks,pdftex]{hyperref}
\idline{94}{1}

\begin{document}
\def\teff{$T\rm_{eff }$}
\def\kms{$\mathrm {km s}^{-1}$}

\title{Climate and atmospheric models of rocky planets: habitability and observational properties}

\subtitle{}

\author{
L. \,Silva\inst{1, 2}
\and R. \, Bevilacqua\inst{3, 1}
\and L. \, Biasiotti\inst{3, 1}
\and E. \, Bisesi\inst{1}
\and S. L. \, Ivanovski\inst{1}
\and M. \, Maris\inst{1, 2}
\and S. \, Monai\inst{1}
\and G. \, Murante\inst{1, 2}
\and P. \, Simonetti\inst{1}
\and G. \, Vladilo\inst{1}
}

\institute{
INAF - National Institute of Astrophysics -
OATs, Via G.B. Tiepolo 11, I-34143 Trieste, Italy
\and
IFPU - Institute for Fundamental Physics of the Universe, Via Beirut 2, I-34014 Trieste, Italy
\and
University of Trieste - Dep. of Physics, Via G.B. Tiepolo 11, I-34143 Trieste, Italy\\
\email{laura.silva@inaf.it}
}

\authorrunning{L. Silva}

\titlerunning{Habitability with climate models}

\date{Received: Day Month Year; Accepted: Day Month Year}

\abstract{The quest for atmospheric spectral signatures that may witness biological activity in exoplanets is focused on rocky planets. The best targets for future, challenging spectroscopic observations will be selected among potentially habitable planets.
Surface habitability can be quantified and explored with climate and atmospheric models according to temperature-based criteria. The conceptual, modellistic, technological and interpretative complexity of the problem requires to develop flexible climate and atmospheric models suited for a comprehensive exploration of observationally unconstrained parameters, and to simulate and interpret definitely non-terrestrial conditions.
We present a summary and preliminary results on the work we are performing on multi-parametric explorations of the habitability and observational properties of rocky planets.
}

\maketitle{}

\section{Introduction}

The surface environmental conditions that may allow a planet to be tagged as potentially habitable is the outcome of a plethora of physical factors involving all scales -- at the planetary, orbital, and stellar level, and continuously acting since the same process of planet formation. 
Therefore both the search for potential habitable candidates as best spectroscopic targets, as well as the ensuing challenging task of interpreting observed data, possibly in terms of biosignatures, rest on thorough theoretical multi-parametric physical modelling of the planetary surfaces and atmospheres. A vast range of unknowns affects also our understanding of the early evolution of Earth, when probably terrestrial life was able to originate, as well as of Mars and Venus, that underwent completely different evolutionary paths. 
We have developed climate and atmospheric modelling tools 
specifically tailored to perform habitability and interpretative studies, to be also applied to simulate future space missions. Here we present a short summary with preliminary results of our work on these subjects.

\section{Climate and atmospheric  model}
\label{sec:models}

In order to cast light on the physical properties and habitability of rocky exoplanets, as well as on the early evolutionary phases of Earth, Mars and Venus, we need to run climate simulations.
The choice of the models depends on how many observational data are available for individual cases. In general, for exoplanets we can measure only a few structural (radius, mass) and orbital (semi-major axis, eccentricity) parameters, plus a few properties of the host star (luminosity, spectral type, chemical composition).
Spectroscopic data of the thin atmospheres of rocky exoplanets are still quite scarce and are limited to planets in tight orbits around very cold stars, conditions that are not particularly favourable for habitability. 
Given the lack of physical/chemical observational constrains on key quantities, climate/atmospheric models used in this field must be able to physically handle them as free parameters, in contrast with models used for the modern Earth, where most parameters are embedded. 
Here we briefly describe a model of this type that has been developed at INAF-Trieste in collaboration with climatologists based at CNR-IGG Pisa and Torino Politecnico.

\subsection{The climate model: ESTM}
\label{sec:ESTM}

The Earth-like planet Surface Temperature Model (ESTM) was developed to quantify the surface planetary habitability starting  from the spatial–temporal distribution of surface temperatures (Vladilo et al. 2015). The core of ESTM is an energy balance model (EBM) with time-dependent quantities, complemented by: radiative–convective atmospheric column calculations, a set of physically based parametrizations of the meridional transport, and descriptions of surface and cloud properties more refined than in classic EBMs. Due to its dependence on time, latitude and vertical atmospheric column, the ESTM is in practice a seasonal, 2D climate model. Thanks to its extremely low computational cost, it can be used to explore the parameter space unconstrained by observations of individual rocky (exo)planets, or to perform statistical studies of climate properties whenever many runs for different parameter configurations are needed. The model has been recently upgraded and its most recent version, EOS-ESTM (Biasiotti et al. 2022), includes a new  code, EOS, (Simonetti et al. 2022; Simonetti 2022) for the treatment of the radiative transfer, as well as a number of parametrizations to simulate the climate impact of oceans, land, ice, and clouds as a function of temperature and host star zenith distance.
EOS-ESTM can be applied to a large variety of rocky planets, with terrestrial and non-terrestrial atmospheric compositions illuminated by solar- and non-solar-type stars. One limitation of this (and any EBM-based) model is that it cannot be applied to tidally-locked planets. 
The model is undergoing a continuous process of cross-validation with a variety of climate models with different levels of complexity.

\subsection{The atmospheric model: EOS}
\label{sec:EOS}

EOS (named after the Greek goddess of Dawn) is a radiative-convective equilibrium model for single atmospheric columns. It is derived from the opacity calculation tool HELIOS-K (Grimm \& Heng 2015; Grimm et al. 2021)  and the radiative transfer code HELIOS (Malik et al. 2017; 2019), of the Exoclime Simulation Platform\footnote{\url{https://github.com/exoclime}}, and has been specifically developed to model the relatively cold and dense atmospheres of terrestrial planets. Special care has been devoted to collision-induced absorption (CIA) continua and to the impact of the moist convection on the tropospheric vertical pressure-temperature profile. As its parent codes, EOS is GPU-accelerated and thus computationally efficient.

EOS operates in the two-stream approximation (see e.g. Pierrehumbert 2010) including non-isotropic scattering. Opacity calculations can be carried on either line-by-line or by using the k-distribution method (Goody \& Yung 1989). The code can take as input any line list or CIA table from popular spectroscopical repositories such as HITRAN (Gordon et al. 2022), HITEMP (Rothman et al. 2010) and ExoMol (Tennyson et al. 2020) and does not depend on hard-wired opacity parameterizations. Similarly, the incoming stellar radiation spectrum can be specified by the user or set to an ideal blackbody of any temperature. The ability to change freely both the atmospheric composition and the instellation  makes EOS extremely flexible and easy to keep updated.

The primary output of EOS is the full reflectance/emission spectrum for a specified set of parameters. From this we get the integrated quantities required by the EBM equation (the Outgoing Longwave Radiation -OLR-, and the top-of-atmosphere -TOA- albedo). 
The  synthetic spectra allows to connect the climate state and habitability, as calculated by ESTM, to observations that will be performed with future instruments (Sec. \ref{sec:mod_obs}).

In practice, radiative transfer calculations for our climate simulations are performed by following the reverse procedure of Kasting (see e.g. Kasting et al. 1993). For any given atmospheric composition we obtain the OLR and the TOA albedo as a function of the surface temperature and reflectivity, the stellar zenith distance and potentially other variables, storing them in lookup tables. These tables can in principle be used also by other climate models, thus being another standalone product of EOS-ESTM.

\section{Model applications}
\label{sec:application}

\subsection{Identifying habitable Exoplanets}

 Most of the currently known exoplanets have been uncovered with indirect detection techniques, i.e. the transit and the radial velocity methods, 
in which the planet's influence on the host star is observed -- luminosity dimming during transits, or Doppler shift of spectral lines due to movement by reciprocal gravitational interaction. Therefore, larger planets in close orbits around small late-type stars, able to determine stronger planet-to-star signals, are preferentially selected (e.g. Winn 2018).
Due to  these observational biases, currently only one potential terrestrial analogue orbiting the habitable zone (HZ) of a G-type star has been detected, Kepler-452b (Jenkins et al. 2015).

In Silva et al. (2017a) we constrained its current and past habitability for different atmospheric pressures and compositions, also accounting for the luminosity evolution of the host star. 
We found that a relatively low CO$_2$ abundance ($\lesssim 0.04$ bar) would allow a current habitable surface. 
Instead, a long-term $>2$ Gyr habitability, that could promote formation and widespread persistence of biological activity 
able to pollute the atmosphere, would have required $\gtrsim 100$ times more greenhouse effect at early times due to stellar evolution effects (as envisaged also for the habitability of the paleo-Earth, Sec. \ref{sec:arch_earth}).

The observational (and statistical) bias against detecting potentially habitable rocky planets is partly tackled for exoplanets orbiting the HZ of M-type stars. Even though the structural planetary properties could be of terrestrial-type (as e.g. the Trappist planetary system), it is expected that the star-planet interaction and evolution determines radically non-terrestrial situations, among which is the possible establishment of the spin-orbit tidal lock. This configuration is not adequate for EBM-type models (but see e.g. Haqq-Misra \& Hayworth 2022). On the other hand, sufficiently eccentric orbits can severely delay or avoid the onset of the synchronous spin-orbit rotation. 
One such case is Gl-514b, a super-Earth recently detected by Damasso et al. (2022). We are investigating with EOS-ESTM the habitability of this planet as a function of a large range of atmospheric properties. Very preliminary calculations show that indeed habitable solutions do exist, with possibly large seasonal excursions (Biasiotti et al., in prep.), that will consequently imply different observational properties. 

\begin{figure}
\centering
\includegraphics[width=0.50\textwidth, height=5.5truecm]{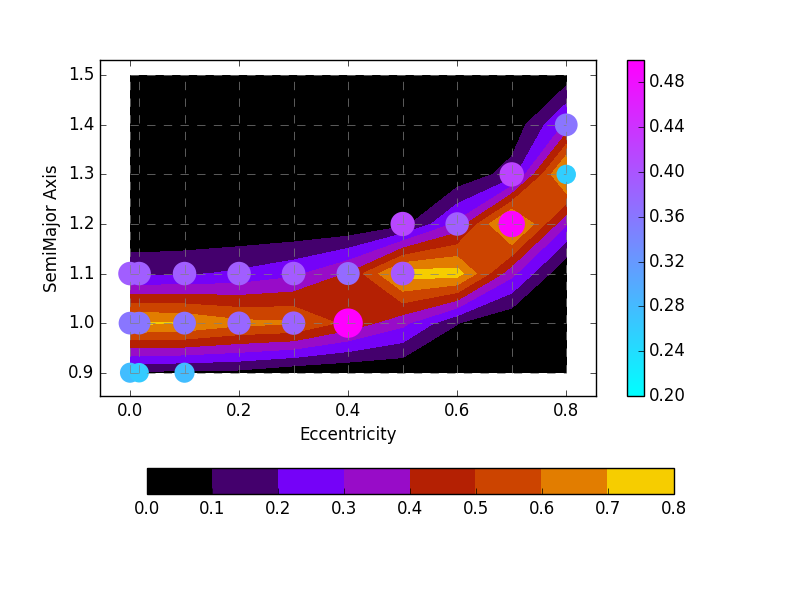}
\caption{ Habitability map on the semi-major orbital axis vs eccentricity plane (color-code in the lower bar; the habitability criterion is the ''complex'' habitability as defined in Silva et al. (2017b). Those solutions admitting climate warm-snowball bistability are shown with filled circles. Their color-code in the right bar gives the snowball probability. Most bistable and habitable solutions are found within the same parameter range. See Murante et al. (2020) for more details.}
\label{fig:bishab}
\end{figure}

A well-known property of the climate system is its chaotic nature, arising from the combined and concomitant action of different non-linear feedbacks, that may give rise to hysteresis cycles and multi-stable solutions. This implies that, as occurred several times to Earth, a planet could be habitable and inhabited even if in a snowball state. This possibility can be accounted for with a probabilistic approach, as we have shown in Murante et al. (2020, M20 hereafter). 

By analyzing a few $\times 10^4$ simulations\footnote{The simulations database, ARTECS, is available at https://wwwuser.oats.inaf.it/exobio/climates/} 
 M20 explored the frequency and parameter dependence of bistable warm-snowball climate solutions. The considered parameters -- pressure, semi-major axis, eccentricity, obliquity -- were coupled to a range of different initial conditions (represented by the initial temperature) to run the simulations. M20 found a probability $\lesssim 10$\% of bistable solutions within the explored range of parameters, but interestingly found an indication of a  connection between habitability and bi-stability, i.e., that the range of parameter values giving rise to bistability is similar to range allowing habitable solutions (Fig. \ref{fig:bishab}).

In addition to the ice-albedo feedback, we have recently introduced in ESTM also the vegetation-albedo feedback (Bisesi et al., in prep.). In fact, the temperature dependence of the vegetation coverage of lands heavily affects the albedo, being significantly smaller for vegetation rather than for 
bare continents ($\sim 0.1$ vs $\sim 0.3$). 
On Earth, the vegetation/albedo feedback is known as Charney mechanism (Charney et al. 1975). It affects the Earth climate and, depending on the planetary properties, it can increase the width of the circumstellar HZ by up to 10\% on its outer border. The Charney mechanism may also increase the global habitability of the planet by a similar amount.

\subsection{The Archean Earth}
\label{sec:arch_earth}

\begin{figure}
\centering
\includegraphics[width=0.49\textwidth]{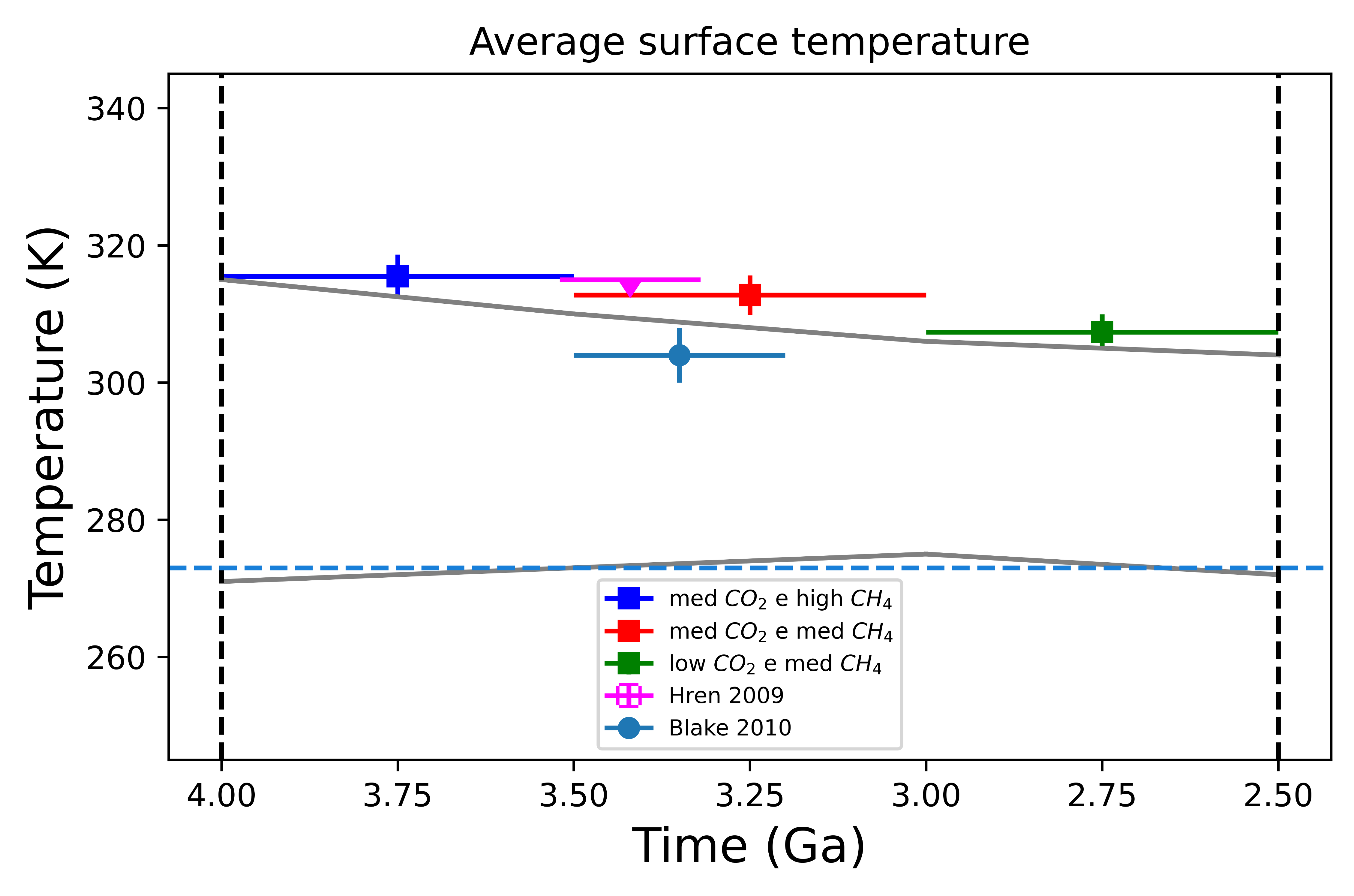}
\caption{Global mean surface temperature of the Earth during the Archean. The colored squares indicate the results obtained in the present work for the 
atmospheric compositions in the legend
(med/low ${CO_2}$ is for p$_{CO_2}$=0.04/0.01 bar; high/med ${CH_4}$ is for 1000/100 ppm). The gray curves are the lower and upper limits of the temperature constraints 
obtained by Catling \& Zahnle (2020), while the magenta upper limit and the cyan circle represent geological constraints on temperature by Hren et al. (2009) and Blake et al. (2010), respectively. The dashed horizontal line is for the water freezing point.}
\label{fig:arch_earth_av_temp}
\end{figure}

The Archean eon spans the time from 4 to 2.5 Gyr ago, i.e. roughly $1/3$ of the Earth evolution, and it probably accommodates the emergence of life. It is therefore particularly important to try to understand the evolution of the environmental conditions ruling habitability during such a long period of time. 
This eon was characterized by an anoxic atmosphere, a shorter rotation period, a very low coverage of continents, and by a fainter insolation ($\sim 0.7-0.8$ the current value, due to stellar evolution). In fact, the well-known Faint Young Sun problem refers to trying to reconcile the geological evidences of the presence of liquid water during the Archean with such a low energy input from the Sun. Among several co-factors, such evidence is generally ascribed to larger amounts of mainly CO$_2$ and/or CH$_4$, able to provide a greenhouse forcing much larger that the current value (i.e. $>30$ $^{\circ}$C). 
According to several geological constraints (e.g. Catling \& Zahnle 2020)
the Earth planetary/atmospheric properties have evolved during the eon, in such a way to maintain on average a temperate surface.     

EOS-ESTM is particularly suited to study the large parameter space that may have allowed habitability of the Archean Earth. We present below some preliminary results that we have performed by considering a range of atmospheric compositions (p$_{CO_2}$=0.01 to 0.1 bar, and ${CH_4}$ abundance = 0 to 1000 ppm, for a total N$_2$-dominated dry pressure of 1 bar with $60\%$ relative humidity), and at the same time by accounting for the time evolution of the luminosity of the Sun (Gough 1981), of the rotation period (Bartlett \& Stevenson 2016), and of the fraction of land coverage (Collerson \& Kamber 1999). For more details see Bevilacqua (2022); Simonetti (2022). 

In Fig. \ref{fig:arch_earth_av_temp} we  show the evolution of the global mean surface temperature estimated from available data and compared to our simulations.
The gray curves are the lower and upper limits of temperatures with 95\% confidence intervals obtained by Catling \& Zahnle (2020). The magenta upper limit and the cyan circle are the constraints from geological data provided respectively by Hren et al. (2009) and Blake et al. (2010).
The squares with error bars are the results obtained from our simulations. 
These are computed as follows. We have split the 1.5 Gyr-length of the eon into three 500 Myr-long intervals. For each step, we selected among our currently computed simulations (that at each age adopt the appropriate insolation and planetary properties) a composition able to provide a habitable surface temperature. 
Specifically: (a) blue square, 4-3.5 Gyr, p$_{CO_2}$=0.04 bar, ${CH_4}$=1000 ppm; (b)  red square, 3.5-3 Gyr, p$_{CO_2}$=0.04 bar, ${CH_4}$=100 ppm; (c) green square, 3-2.5 Gyr, p$_{CO_2}$=0.01 bar, ${CH_4}$=100 ppm.
The temperatures we obtained are consistent with the geological constraints derived by Hren et al. (2009) and Blake et al. (2010). 

In each phase, by considering the Earth as a benchmark for exoplanets, the observable spectral features would be different. 
We are working to produce a finer model grid in order to explore a larger range of possible solutions, and also to simulate for each case, the corresponding expected observable atmospheric spectrum.

\section{Connecting models of climate and habitability to observations}
\label{sec:mod_obs}

\begin{figure*}
\centering
\includegraphics[width=0.32\textwidth]{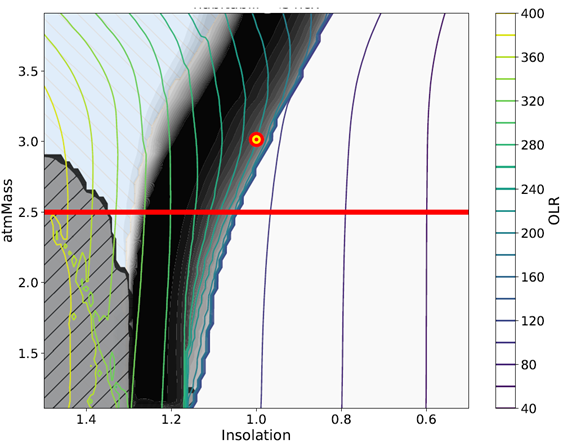}
\includegraphics[width=0.32\textwidth]{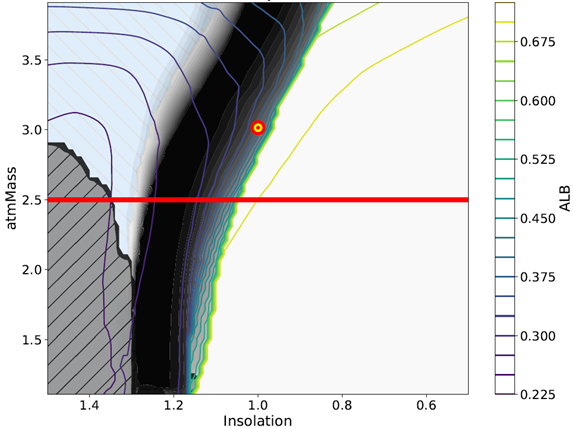}
\includegraphics[width=0.32\textwidth]{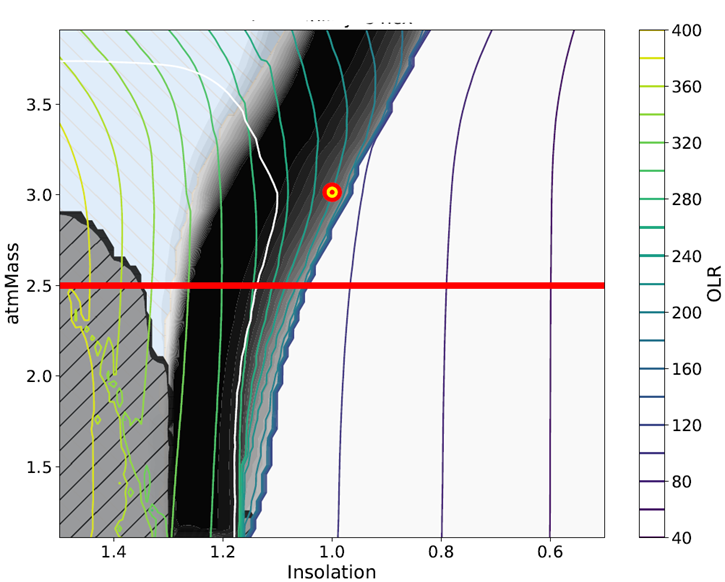}
\caption{Habitability maps in the atmospheric mass ($\log p/g$ in gr/cm$^2$) vs insolation plane, as in Silva et al. (2017b), with superimposed the contour lines of constant OLR (left panel) and TOA albedo (central panel). These curves decouple and intesect at large enough $p/g$ (right panel, where for clarity only one of the albedo lines is shown in white together the OLR lines). 
Since OLR and TOA albedo, together with insolation, are observables, they could provide a measure of the habitability.
The red-yellow circle is the position of the Earth. The thick red line marks the atmospheric mass limit below which the surface radiation dose of secondary particles of Galactic Cosmic Rays would be $>100$ mSv/yr (see Silva et al. 2017b).}
\label{fig:obs_hab}
\end{figure*}

The atmospheric characterization of rocky exoplanets may be hopefully achievable already during the JWST lifetime for M-type hosts, but it is considered a fundamental target for already planned or under design space- and ground-based instruments during the next decades (e.g. ARIEL, ELT, LUVOIR, HabEX, LIFE). 
Depending on the line-of-sights configurations, orbital distances and wavelengths, spectra will be collected via   transmission, reflection or emission.
The same atmospheric structure and composition that dictates the climate and the surface environmental conditions for habitability\footnote{Habitability can be quantified with different possible temperature-based criteria once the climate state has been computed. Also, it can be extended to include the protective role of the atmosphere against ionizing radiation (see Silva et al. 2017b).}, 
rules also the spectral emission that will be measured. It is therefore necessary to model at the same time climate and habitability, with the associated atmospheric spectrum.

As an example of the connection between observables and habitability, in Fig. \ref{fig:obs_hab} we show the habitability map as a function of atmospheric mass ($p/g$) and insolation, as in Silva et al. (2017b), with superimposed the corresponding contours of constant OLR (left panel), and TOA albedo (central panel). At large enough $p/g$ values, these contour lines decouple and intersect. The observational evaluation of OLR and Bond albedo, together with the measured insolation, could therefore in principle provide an estimate of the habitability. 
This sort of statistical inferences requires large numbers of model simulations, which we can perform with our modelling tools.

\subsection{Transmission spectra: effect of refraction}
\label{sec:refraction}

\begin{figure*}
\centering
\includegraphics[width=0.49\textwidth]{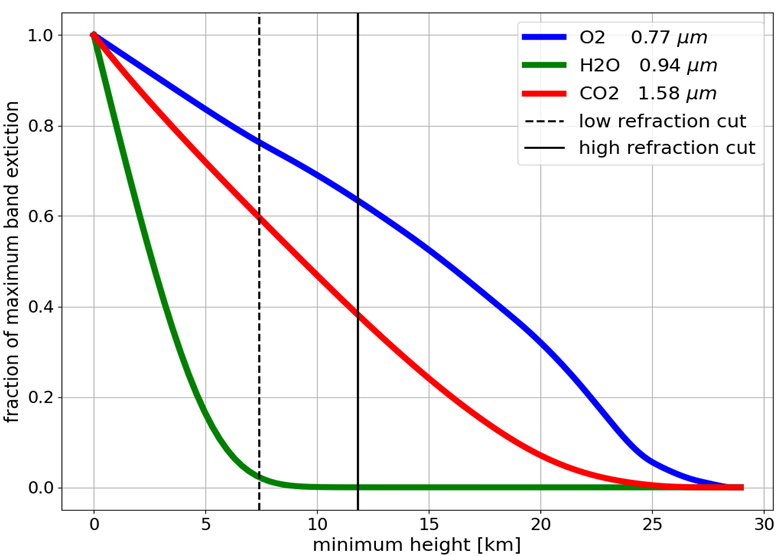}
\includegraphics[width=0.49\textwidth, height=4.7truecm]{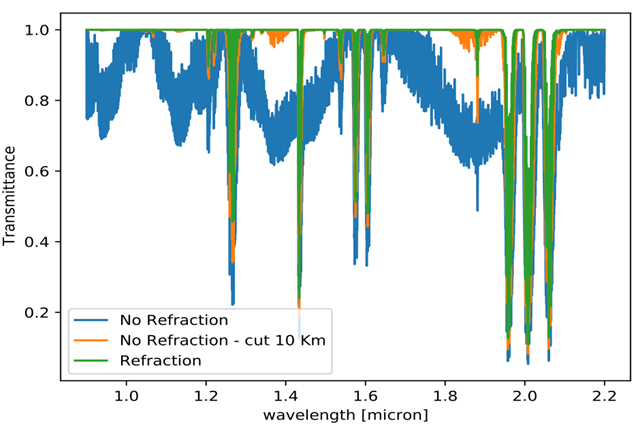}
\caption{{\it Left}: Fraction of observed integrated band extinction with respect to no refraction, vs the minimum probed atmospheric height, for three representative bands for O$_2$ ($0.759-0.775$ $\mu$m), CO$_2$ ($1.565-1.591$ $\mu$m), H$_2$O ($0.860-1.020$  $\mu$m). This example is for a current Earth-like atmosphere
in the best case of clear-sky conditions. 
The vertical dashed and solid lines show the minimum heights probed at transit start and center, respectively ($7.4$ and $11.8$ km for this model). All layers above the latter minimum height are always visible during transit. {\it Right}: transmittance spectrum of molecular absorption in the $\sim 1-2 \, \mu$m region. The blue and orange lines are for no refraction, but in the latter case a 10\,km cut is applied to represent a 100\% cloud coverage. This case is almost superimposed to the green line, that is the transmittance that is obtained including refraction. See text for more details. From Maris et al. (in prep).}
\label{fig:refraction}
\end{figure*}

Exoplanets discovered with the transit method -- so far the majority -- are in principle good candidates for measuring their atmospheric spectrum in transmission. However, transmission spectra are affected by refraction, which impacts the accessible atmospheric layers. 
The deflection angles depend on atmospheric composition and density, and on the angular size of the host star as seen from the planet.
Therefore refraction is an issue for the relatively dense atmospheres of terrestrial-type rather than giant planets, and in particular for habitable rocky planets around G-type stars due to the implied geometry.    

We present preliminary results from Maris et al. (in prep.) in simulating observations that account for refraction. We highlight that the same atmospheric models that, with EOS-ESTM, determine the climate state, are then used to compute the associated refraction and observability of the spectral features.

Fig.\ \ref{fig:refraction}, left panel,  
shows the fraction, with respect to no refraction, of the integrated band extinction for 3 representative bands for O$_2$ ($0.759-0.775$ $\mu$m), CO$_2$ ($1.565-1.591$ $\mu$m) and H$_2$O ($0.860-1.020$  $\mu$m), that would be observed, for a model of the modern Earth, at different sampled atmospheric heights. For this model, only the layers above $11.8$ km (i.e. in practice above almost all the troposhere containing most atmospheric mass) would be always visible during transit, implying that the depths of the considered bands would be reduced to $\sim 60$\% for O$_2$ and 40\% for CO$_2$ compared to their full depth. Instead the water vapor band would be totally missed. Note that the shown example is for the best ideal clear-sky case.
In the right panel we show an illustrative simplified example of the possible effects of clouds on the transmittance of the molecular absorption spectra.
The blue and orange spectra are without refraction, but for the latter spectrum a 10 km cut has been applied in the sampled atmospheric layers, 
in order to schematically represent a 100\% cloud coverage. This case is almost superimposed to the green line, the transmittance that is obtained accounting for refraction. 
Note that the shown examples do not include the Rayleigh scattering. This and other effects and complications will be presented in Maris et al. (in prep.). 

While clouds (and hazes) will be a severe issue for the detection of spectral features and biosignatures also for habitable planets in close orbits around M-stars hosts, refraction will severely limit the possibility of detecting features precisely for habitable terrestrial analogues in any case. In these situations, emission spectra could provide better probes.

\subsection{Emission spectra}
\label{sec:emission}

Emission/reflection spectra are valuable tools that both allow for the characterization of the disk-integrated vertical pressure-temperature structure of the atmosphere and give information on the horizontal heat distribution. An example of this type of analysis has been conducted on WASP-103b (Kreidberg et al. 2018). Currently, a few low resolution emission/reflection spectra of hot and ultra-hot Jupiters are available, while proposed future missions such as the Large Interferometer For Exoplanets (LIFE, Quanz et al. 2022) will focus on the study of rocky, temperate planets.

\begin{figure}
\centering
\includegraphics[width=0.5\textwidth]{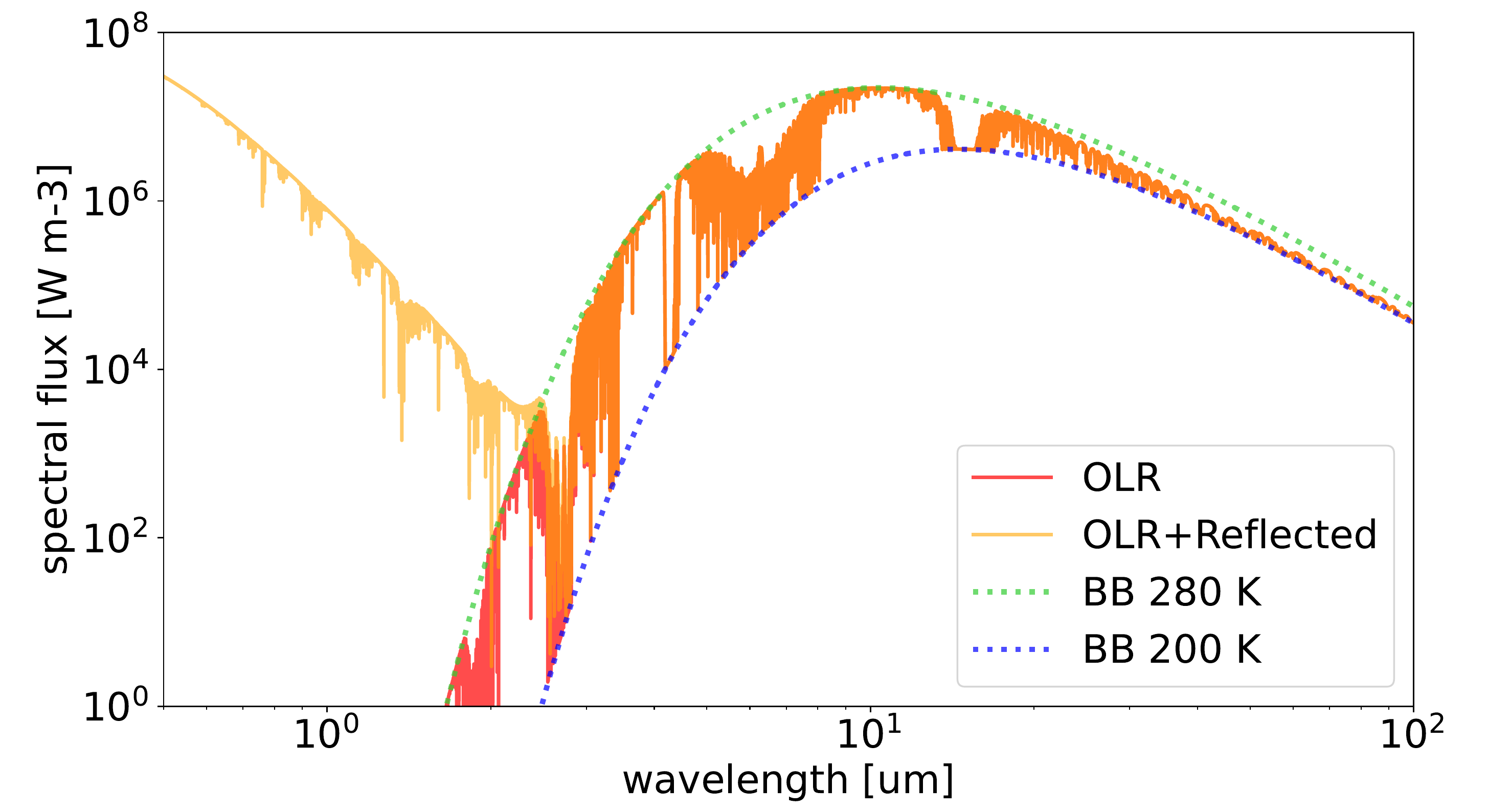}
\caption{Top of atmosphere SED for an Earth-like planet with a surface temperature of 280 K, a tropopause at 200 K and a surface reflectance of 0. The instellation is produced by a blackbody at 5780 K, emulating the Sun. The red line represents the OLR only, while the yellow line represents the sum of OLR and the reflected stellar radiation due to atmospheric Rayleigh scattering. The dotted lines describes the blackbody curves for a 280 K and a 200 K emitter.}
\label{fig:earth_spec}
\end{figure}

As an intermediate step in producing the OLR and TOA albedo lookup tables, EOS calculates the synthetic emission/reflection spectrum of the planet. An example of this is shown in Fig.~\ref{fig:earth_spec}, which has been produced using an Earth-like atmosphere containing N$_2$, O$_2$, 360 ppm of CO$_2$, 1.8 ppm of CH$_4$ and a varying amount of H$_2$O, calculated fixing the relative humidity to 60\%. The surface temperature has been fixed to 280 K and the vertical pressure-temperature profile has been specified in input, using a moist adiabatic lapse rate for the troposphere and a 200 K isothermal stratosphere. The surface albedo has been set to 0 and the incoming stellar radiation has been produced by a blackbody at 5780 K, with a total flux normalized to the Earth's Solar Constant (1361 W m$^{-2}$). The calculation has been carried on in the $[0.34,200]$ $\mu$m range at a $\lambda/\Delta\lambda$ resolution of 3000. 
This example, calculated for a single atmospheric column, shows that, with the disc-integrated emission spectrum appropriate to each EOS-ESTM simulation, we will be able to test the detectability of habitable exoplanets with future space missions  (e.g. LIFE).

\begin{acknowledgements}
The research reported in this work was supported by the 
Italian Space Agency with the Life in Space project (ASI N. 2019-3-U.0) and by OGS and CINECA under HPC-TRES program award number 2022-02.
\end{acknowledgements}



\begin{thebibliography}{}
	
	\expandafter\ifx\csname natexlab\endcsname\relax\def\natexlab#1{#1}\fi
	\providecommand{\url}[1]{\href{#1}{#1}}
	\providecommand{\dodoi}[1]{doi:~\href{http://doi.org/#1}{\nolinkurl{#1}}}
	\providecommand{\doeprint}[1]{\href{http://ascl.net/#1}{\nolinkurl{http://ascl.net/#1}}}
	\providecommand{\doarXiv}[1]{\href{https://arxiv.org/abs/#1}{\nolinkurl{https://arxiv.org/abs/#1}}}
	
	\bibitem[{{Bartlett} \& {Stevenson}(2016)}]{bartlett16}
	{Bartlett}, B.~C., \& {Stevenson}, D.~J. 2016, \grl, 43, 5716,
	\dodoi{10.1002/2016GL068912}
	
	\bibitem[{{Bevilacqua}(2022)}]{bevilacqua22}
	{Bevilacqua}, R. 2022, Master Thesis, Universit\`a degli Studi di Trieste.
	\newblock \url{http://hdl.handle.net/20.500.12072/92838}
	
	\bibitem[{{Biasiotti} {et~al.}(2022){Biasiotti}, {Simonetti}, {Vladilo},
		{Silva}, {Murante}, {Ivanovski}, {Maris}, {Monai}, {Bisesi}, {von
			Hardenberg}, \& {Provenzale}}]{bia22}
	{Biasiotti}, L., {Simonetti}, P., {Vladilo}, G., {et~al.} 2022, \mnras, 514,
	5105, \dodoi{10.1093/mnras/stac1642}
	
	\bibitem[{{Blake} {et~al.}(2010){Blake}, {Chang}, \& {Lepland}}]{blake2010}
	{Blake}, R.~E., {Chang}, S.~J., \& {Lepland}, A. 2010, \nat, 464, 1029,
	\dodoi{10.1038/nature08952}
	
	\bibitem[{{Catling} \& {Zahnle}(2020)}]{catzah20}
	{Catling}, D.~C., \& {Zahnle}, K.~J. 2020, Science Advances, 6, eaax1420,
	\dodoi{10.1126/sciadv.aax1420}
	
	\bibitem[{Charney {et~al.}(1975)Charney, Stone, \& Quirk}]{charney75}
	Charney, J., Stone, P.~H., \& Quirk, W.~J. 1975, Science, 187, 434,
	\dodoi{10.1126/science.187.4175.434}
	
	\bibitem[{{Collerson} \& {Kamber}(1999)}]{collerson99}
	{Collerson}, K.~D., \& {Kamber}, B.~S. 1999, Science, 283, 1519,
	\dodoi{10.1126/science.283.5407.1519}
	
	\bibitem[{{Damasso} {et~al.}(2022){Damasso}, {Perger}, {Almenara}, {Nardiello},
		{P{\'e}rez-Torres}, {Sozzetti}, {Hara}, {Quirrenbach}, {Bonfils}, {Zapatero
			Osorio}, {Astudillo-Defru}, {Gonz{\'a}lez Hern{\'a}ndez}, {Su{\'a}rez
			Mascareno}, {Amado}, {Forveille}, {Lillo-Box}, {Alibert}, {Caballero},
		{Cifuentes}, {Delfosse}, {Figueira}, {Galad{\'\i}-Enr{\'\i}quez}, {Hatzes},
		{Henning}, {Kaminski}, {Mayor}, {Murgas}, {Montes}, {Pinamonti}, {Reiners},
		{Ribas}, {B{\'e}jar}, {Schweitzer}, \& {Zechmeister}}]{damasso22}
	{Damasso}, M., {Perger}, M., {Almenara}, J.~M., {et~al.} 2022, \aap, 666, A187,
	\dodoi{10.1051/0004-6361/202243522}
	
	\bibitem[{{Goody} \& {Yung}(1989)}]{goody89}
	{Goody}, R.~M., \& {Yung}, Y.~L. 1989, {Atmospheric radiation : theoretical
		basis} (New York, USA: Oxford University Press)
	
	\bibitem[{{Gordon} {et~al.}(2022){Gordon}, {Rothman}, {Hargreaves}, {Hashemi},
		{Karlovets}, {Skinner}, {Conway}, {Hill}, {Kochanov}, {Tan}, {Wcis{\l}o},
		{Finenko}, {Nelson}, {Bernath}, {Birk}, {Boudon}, {Campargue}, {Chance},
		{Coustenis}, {Drouin}, {Flaud}, {Gamache}, {Hodges}, {Jacquemart}, {Mlawer},
		{Nikitin}, {Perevalov}, {Rotger}, {Tennyson}, {Toon}, {Tran}, {Tyuterev},
		{Adkins}, {Baker}, {Barbe}, {Can{\`e}}, {Cs{\'a}sz{\'a}r}, {Dudaryonok},
		{Egorov}, {Fleisher}, {Fleurbaey}, {Foltynowicz}, {Furtenbacher}, {Harrison},
		{Hartmann}, {Horneman}, {Huang}, {Karman}, {Karns}, {Kassi}, {Kleiner},
		{Kofman}, {Kwabia-Tchana}, {Lavrentieva}, {Lee}, {Long}, {Lukashevskaya},
		{Lyulin}, {Makhnev}, {Matt}, {Massie}, {Melosso}, {Mikhailenko}, {Mondelain},
		{M{\"u}ller}, {Naumenko}, {Perrin}, {Polyansky}, {Raddaoui}, {Raston},
		{Reed}, {Rey}, {Richard}, {T{\'o}bi{\'a}s}, {Sadiek}, {Schwenke},
		{Starikova}, {Sung}, {Tamassia}, {Tashkun}, {Vander Auwera}, {Vasilenko},
		{Vigasin}, {Villanueva}, {Vispoel}, {Wagner}, {Yachmenev}, \&
		{Yurchenko}}]{gordon22}
	{Gordon}, I.~E., {Rothman}, L.~S., {Hargreaves}, R.~J., {et~al.} 2022, \jqsrt,
	277, 107949, \dodoi{10.1016/j.jqsrt.2021.107949}
	
	\bibitem[{{Gough}(1981)}]{gough81}
	{Gough}, D.~O. 1981, \solphys, 74, 21, \dodoi{10.1007/BF00151270}
	
	\bibitem[{{Grimm} \& {Heng}(2015)}]{grimm15}
	{Grimm}, S.~L., \& {Heng}, K. 2015, \apj, 808, 182,
	\dodoi{10.1088/0004-637X/808/2/182}
	
	\bibitem[{{Grimm} {et~al.}(2021){Grimm}, {Malik}, {Kitzmann},
		{Guzm{\'a}n-Mesa}, {Hoeijmakers}, {Fisher}, {Mendon{\c{c}}a}, {Yurchenko},
		{Tennyson}, {Alesina}, {Buchschacher}, {Burnier}, {Segransan}, {Kurucz}, \&
		{Heng}}]{grimm21}
	{Grimm}, S.~L., {Malik}, M., {Kitzmann}, D., {et~al.} 2021, \apjs, 253, 30,
	\dodoi{10.3847/1538-4365/abd773}
	
	\bibitem[{{Haqq-Misra} \& {Hayworth}(2022)}]{haqq22}
	{Haqq-Misra}, J., \& {Hayworth}, B. P.~C. 2022, PSJ, 3, 32,
	\dodoi{10.3847/PSJ/ac49eb}
	
	\bibitem[{{Hren} {et~al.}(2009){Hren}, {Tice}, \& {Chamberlain}}]{hren2009}
	{Hren}, M.~T., {Tice}, M.~M., \& {Chamberlain}, C.~P. 2009, \nat, 462, 205,
	\dodoi{10.1038/nature08518}
	
	\bibitem[{{Jenkins} {et~al.}(2015){Jenkins}, {Twicken}, {Batalha}, {Caldwell},
		{Cochran}, {Endl}, {Latham}, {Esquerdo}, {Seader}, {Bieryla}, {Petigura},
		{Ciardi}, {Marcy}, {Isaacson}, {Huber}, {Rowe}, {Torres}, {Bryson},
		{Buchhave}, {Ramirez}, {Wolfgang}, {Li}, {Campbell}, {Tenenbaum},
		{Sanderfer}, {Henze}, {Catanzarite}, {Gilliland}, \& {Borucki}}]{jenkins15}
	{Jenkins}, J.~M., {Twicken}, J.~D., {Batalha}, N.~M., {et~al.} 2015, \aj, 150,
	56, \dodoi{10.1088/0004-6256/150/2/56}
	
	\bibitem[{{Kasting} {et~al.}(1993){Kasting}, {Whitmire}, \&
		{Reynolds}}]{kasting93}
	{Kasting}, J.~F., {Whitmire}, D.~P., \& {Reynolds}, R.~T. 1993, Icarus, 101,
	108, \dodoi{10.1006/icar.1993.1010}
	
	\bibitem[{{Kreidberg} {et~al.}(2018){Kreidberg}, {Line}, {Parmentier},
		{Stevenson}, {Louden}, {Bonnefoy}, {Faherty}, {Henry}, {Williamson},
		{Stassun}, {Beatty}, {Bean}, {Fortney}, {Showman}, {D{\'e}sert}, \&
		{Arcangeli}}]{kreidberg18}
	{Kreidberg}, L., {Line}, M.~R., {Parmentier}, V., {et~al.} 2018, \aj, 156, 17,
	\dodoi{10.3847/1538-3881/aac3df}
	
	\bibitem[{{Malik} {et~al.}(2019){Malik}, {Kitzmann}, {Mendon{\c{c}}a}, {Grimm},
		{Marleau}, {Linder}, {Tsai}, \& {Heng}}]{malik19}
	{Malik}, M., {Kitzmann}, D., {Mendon{\c{c}}a}, J.~M., {et~al.} 2019, \aj, 157,
	170, \dodoi{10.3847/1538-3881/ab1084}
	
	\bibitem[{{Malik} {et~al.}(2017){Malik}, {Grosheintz}, {Mendon{\c{c}}a},
		{Grimm}, {Lavie}, {Kitzmann}, {Tsai}, {Burrows}, {Kreidberg}, {Bedell},
		{Bean}, {Stevenson}, \& {Heng}}]{malik17}
	{Malik}, M., {Grosheintz}, L., {Mendon{\c{c}}a}, J.~M., {et~al.} 2017, \aj,
	153, 56, \dodoi{10.3847/1538-3881/153/2/56}
	
	\bibitem[{{Murante} {et~al.}(2020){Murante}, {Provenzale}, {Vladilo},
		{Taffoni}, {Silva}, {Palazzi}, {Hardenberg}, {Maris}, {Londero}, {Knapic}, \&
		{Zorba}}]{murante20}
	{Murante}, G., {Provenzale}, A., {Vladilo}, G., {et~al.} 2020, \mnras, 492,
	2638, \dodoi{10.1093/mnras/stz3529}
	
	\bibitem[{{Pierrehumbert}(2010)}]{pierrehumbert10}
	{Pierrehumbert}, R.~T. 2010, {Principles of Planetary Climate} (Cambridge, UK:
	Cambridge University Press)
	
	\bibitem[{{Quanz} {et~al.}(2022){Quanz}, {Ottiger}, {Fontanet}, {Kammerer},
		{Menti}, {Dannert}, {Gheorghe}, {Absil}, {Airapetian}, {Alei}, {Allart},
		{Angerhausen}, {Blumenthal}, {Buchhave}, {Cabrera},
		{Carri{\'o}n-Gonz{\'a}lez}, {Chauvin}, {Danchi}, {Dandumont}, {Defr{\'e}re},
		{Dorn}, {Ehrenreich}, {Ertel}, {Fridlund}, {Garc{\'\i}a Mu{\~n}oz},
		{Gasc{\'o}n}, {Girard}, {Glauser}, {Grenfell}, {Guidi}, {Hagelberg},
		{Helled}, {Ireland}, {Janson}, {Kopparapu}, {Korth}, {Kozakis}, {Kraus},
		{L{\'e}ger}, {Leedj{\"a}rv}, {Lichtenberg}, {Lillo-Box}, {Linz}, {Liseau},
		{Loicq}, {Mahendra}, {Malbet}, {Mathew}, {Mennesson}, {Meyer}, {Mishra},
		{Molaverdikhani}, {Noack}, {Oza}, {Pall{\'e}}, {Parviainen}, {Quirrenbach},
		{Rauer}, {Ribas}, {Rice}, {Romagnolo}, {Rugheimer}, {Schwieterman},
		{Serabyn}, {Sharma}, {Stassun}, {Szul{\'a}gyi}, {Wang}, {Wunderlich},
		{Wyatt}, \& {LIFE Collaboration}}]{quanz22}
	{Quanz}, S.~P., {Ottiger}, M., {Fontanet}, E., {et~al.} 2022, \aap, 664, A21,
	\dodoi{10.1051/0004-6361/202140366}
	
	\bibitem[{{Rothman} {et~al.}(2010){Rothman}, {Gordon}, {Barber}, {Dothe},
		{Gamache}, {Goldman}, {Perevalov}, {Tashkun}, \& {Tennyson}}]{rothman10}
	{Rothman}, L.~S., {Gordon}, I.~E., {Barber}, R.~J., {et~al.} 2010, \jqsrt, 111,
	2139, \dodoi{10.1016/j.jqsrt.2010.05.001}
	
	\bibitem[{{Silva} {et~al.}(2017{\natexlab{a}}){Silva}, {Vladilo}, {Murante}, \&
		{Provenzale}}]{silva17b}
	{Silva}, L., {Vladilo}, G., {Murante}, G., \& {Provenzale}, A.
	2017{\natexlab{a}}, \mnras, 470, 2270, \dodoi{10.1093/mnras/stx1396}
	
	\bibitem[{{Silva} {et~al.}(2017{\natexlab{b}}){Silva}, {Vladilo}, {Schulte},
		{Murante}, \& {Provenzale}}]{silva17a}
	{Silva}, L., {Vladilo}, G., {Schulte}, P.~M., {Murante}, G., \& {Provenzale},
	A. 2017{\natexlab{b}}, International Journal of Astrobiology, 16, 244,
	\dodoi{10.1017/S1473550416000215}
	
	\bibitem[{{Simonetti}(2022)}]{sim22phd}
	{Simonetti}, P. 2022, PhD thesis, Universit\`a degli Studi di Trieste.
	\newblock \url{https://hdl.handle.net/11368/3030923}
	
	\bibitem[{{Simonetti} {et~al.}(2022){Simonetti}, {Vladilo}, {Silva}, {Maris},
		{Ivanovski}, {Biasiotti}, {Malik}, \& {von Hardenberg}}]{sim22}
	{Simonetti}, P., {Vladilo}, G., {Silva}, L., {et~al.} 2022, \apj, 925, 105,
	\dodoi{10.3847/1538-4357/ac32ca}
	
	\bibitem[{{Tennyson} {et~al.}(2020){Tennyson}, {Yurchenko}, {Al-Refaie},
		{Clark}, {Chubb}, {Conway}, {Dewan}, {Gorman}, {Hill}, {Lynas-Gray},
		{Mellor}, {McKemmish}, {Owens}, {Polyansky}, {Semenov}, {Somogyi}, {Tinetti},
		{Upadhyay}, {Waldmann}, {Wang}, {Wright}, \& {Yurchenko}}]{tennyson20}
	{Tennyson}, J., {Yurchenko}, S.~N., {Al-Refaie}, A.~F., {et~al.} 2020, \jqsrt,
	255, 107228, \dodoi{10.1016/j.jqsrt.2020.107228}
	
	\bibitem[{{Vladilo} {et~al.}(2015){Vladilo}, {Silva}, {Murante}, {Filippi}, \&
		{Provenzale}}]{vla15}
	{Vladilo}, G., {Silva}, L., {Murante}, G., {Filippi}, L., \& {Provenzale}, A.
	2015, \apj, 804, 50, \dodoi{10.1088/0004-637X/804/1/50}
	
	\bibitem[{{Winn}(2018)}]{winn18}
	{Winn}, J.~N. 2018, in Handbook of Exoplanets, ed. H.~J. {Deeg} \& J.~A.
	{Belmonte}, 195, \dodoi{10.1007/978-3-319-55333-7_195}
	
\end{thebibliography}

\end{document}